# Electronic reconstruction through the structural and magnetic transitions in detwinned NaFeAs

**M Yi[1,2], D H Lu[3], R G Moore[1], K Kihou[4,5], C-H Lee[4,5], A Iyo[4,5], H Eisaki[4,5], T Yoshida[5,6], A Fujimori[5,6], Z-X Shen[1,2]***

[1]Stanford Institute for Materials and Energy Sciences, SLAC National Accelerator Laboratory, 2575 Sand Hill Road, Menlo Park, California 94025, USA

[2]Geballe Laboratory for Advanced Materials, Departments of Physics and Applied Physics, Stanford University, Stanford, California 94305, USA

[3]Stanford Synchrotron Radiation Lightsource, SLAC National Accelerator Laboratory, 2575 Sand Hill Road, Menlo Park, California 94025, USA

[4]National Institute of Advanced Industrial Science and Technology, Tsukuba, Ibaraki 305-8568, Japan

[5]JST, Transformative Research-Project on Iron Pnictides, Tokyo, 102-0075, Japan

[6]Department of Physics and Department of Complexity Science and Engineering, University of Tokyo, Hongo, Tokyo 113-0033, Japan

*corresponding author: zxshen@stanford.edu

PACS numbers: 74.25.Jb, 74.70.Xa, 79.60.-i

**Abstract.** We use angle-resolved photoemission spectroscopy to study twinned and detwinned iron pnictide compound NaFeAs. Distinct signatures of electronic reconstruction are observed to occur at the structural ($T_S$) and magnetic ($T_{SDW}$) transitions. At $T_S$, $C_4$ rotational symmetry is broken in the form of an anisotropic shift of the orthogonal $d_{xz}$ and $d_{yz}$ bands. The magnitude of this orbital anisotropy rapidly develops to near completion upon approaching $T_{SDW}$, at which temperature band folding occurs via the antiferromagnetic ordering wave vector. Interestingly, the anisotropic band shift onsetting at $T_S$ develops in such a way to enhance the nesting conditions in the $C_2$ symmetric state, hence is intimately correlated with the long range collinear AFM order. Furthermore, the similar behaviors of the electronic reconstruction in NaFeAs and $Ba(Fe_{1-x}Co_x)_2As_2$ suggests that this rapid development of large orbital anisotropy between $T_S$ and $T_{SDW}$ is likely a general feature of the electronic nematic phase in the iron pnictides, and the associated orbital fluctuations may play an important role in determining the ground state properties.

## 1. Introduction

Among the expansive galore of discovered iron pnictide materials up to date, all those families capable of achieving high $T_C$ possess in the underdoped region of their phase diagrams a collinear spin density wave (SDW) ordered phase whose ordering temperature is closely correlated with a tetragonal to orthorhombic structural transition [1,2]. While the orthorhombic distortion associated with the structural transition breaks the rotational symmetry, the collinear antiferromagnetic (AFM) spin structure formed upon the magnetic transition breaks the translational symmetry. The symmetry-breaking nature of these phases in proximity to high temperature superconductivity in the pnictides makes the study of these competing phases one of high current interest, especially in light of evidences for various broken-symmetries in the pseudogap phase in proximity to superconductivity in the high $T_C$ cuprates [3-5].

Recently, the success of applying uniaxial pressure to detwin pnictide crystals in the orthorhombic phase have led to observations of large in-plane anisotropy in resistivity [6-9] and optical conductivity [10-11]. Similar methods also allowed angle-resolved photoemission spectroscopy (ARPES) studies to observe intrinsic broken symmetry of the underlying electronic structure in the SDW state of detwinned $BaFe_2As_2$ [12,13]. Moreover, our previous study on $Ba(Fe_{1-x}Co_x)_2As_2$ (Ba122) revealed that the electronic anisotropy is also manifested in the orbital degree of freedom as an unequal occupation of the $d_{xz}$ and $d_{yz}$ orbitals. This anisotropy onsets near $T_S$, which occurs above $T_{SDW}$ for x>0, evidencing a non-trivial electronic nematic state in the Ba122 compounds [12].

Up to date, most studies of electronic nematicity in the pnictides have been limited to the 122 compounds. As the magnetic and structural transitions are common features in the phase diagrams of most pnictide families, it is important to establish whether the orbital anisotropy observed in Ba122 is a fundamental phenomenon associated with these competing phases across pnictide families. In this work, we use ARPES to study the evolution of electronic structure through the structural and SDW transitions in NaFeAs, which is the parent compound of a 111 family [16,17]. NaFeAs is an ideal system for ARPES studies due to presence of a charge neutral cleavage plane that minimizes surface problems, as have been reported for both 1111 [16-18] and 122 [19] compounds. Moreover, the large separation of the structural transition ($T_S$=53K) and SDW transition ($T_{SDW}$=40K) allows clear identification of the evolution of electronic structure through each transition. Previous ARPES study on twinned NaFeAs [20] reported a rapid shifting of bands onsetting at $T_S$, which was attributed to band folding due to SDW ordering. In this study, our measurements on detwinned crystals shows that the rapid band shift is actually a symmetry breaking process in the orbital degree of freedom similar to that observed in Ba122, where the bands of dominant $d_{yz}$ orbital character shift up while bands of dominant $d_{xz}$ orbital character shift down in the orthogonal direction, breaking the $C_4$ rotational symmetry. Furthermore, comparison between measurements on detwinned and strain-free twinned crystals shows that this rapid symmetry breaking process onsets near $T_S$ in a strain-free crystal, and the magnitude of orbital anisotropy is almost complete at $T_{SDW}$, resulting in better nested Fermi surfaces (FS) compared to above $T_S$. Subsequently, clear signatures of band folding in this symmetry-broken state are observed to onset at $T_{SDW}$. The clean case of detwinned NaFeAs presented here clarifies the question of how underlying electronic structure evolves through each of the competing phases in iron pnictides as previously complicated by extra surface features in 1111 and 122 as well as those due to domain mixing; and in conjunction with results from 122, also suggests that the electronic nematicity is likely a fundamental phenomenon in underdoped pnictides.

## 2.    Experimental

High quality NaFeAs single crystals were grown by flux method following the recipe described in Kihou *et al* [20], where NaAs was used as a flux. Sample preparation was done in nitrogen glove box to minimize air exposure. ARPES measurements were carried out at beamline 5-4 of the Stanford Synchrotron Radiation Lightsource using a SCIENTA R4000 electron analyzer. ARPES spectra were recorded using 25 eV photons. The total energy resolution was set to 12 meV or better and the angular resolution was 0.3°. Single crystals were cleaved in situ at 10 K for all measurements. All measurements were done in ultra high vacuum chambers with a base pressure better than $4 \times 10^{-11}$ torr. Crystal detwinning was done similarly as previously reported [12]. Brillouin zone (BZ) notations pertaining to the 2-Fe unit cell is used throughout.

## 3.    Results

### 3.1. Electronic structure in the paramagnetic state

The observed electronic structure of NaFeAs in the tetragonal state (Fig.1) is similar to other pnictides, with three hole bands near $\Gamma$ and two electron bands near X. However, one of the hole bands does not appear to cross the Fermi level ($E_F$), leaving 2 hole pockets and 2 electron pockets on the FS (Fig. 1a), as previously reported [21]. Comparison with LDA calculations [22] shows that an overall renormalization factor of 4 produces a rough match with the overall bandwidths of measured dispersions near $E_F$ (Fig. 1b), indicating that correlation strength may be stronger in NaFeAs than that in Ba122 [23]. Moreover, a careful look shows that a momentum-dependent shift fine tune is still needed as the LDA bands are higher than measured at $\Gamma$, but lower than measured at X, similar to Ba122 [23].

Information of the dominant orbital characters can also be inferred from the analysis of light polarization and orbital symmetries [24], in conjunction with comparison with band structure calculations. According to calculations [25], only the $d_{xy}$, $d_{xz}$, and $d_{yz}$ orbitals contribute significantly to the bands near $E_F$. Along $\Gamma$-X in the paramagnetic (PM) state, the $\alpha$, $\beta$, and $\gamma$ hole bands are predicted to be dominantly $d_{xz}$, $d_{yz}$, and $d_{xy}$, respectively, and the $\delta$ and $\varepsilon$ electron bands dominantly $d_{xz}$ and $d_{xy}$. These assignments are consistent with the symmetry constrains in our experiments. Symmetry considerations for the tetragonal state also require $d_{xz}$ and $d_{yz}$ orbitals to be degenerate at the $\Gamma$ point in the PM state, which appears to be at odds with the assignment of $d_{xz}$ and $d_{yz}$ character to $\alpha$ and $\beta$ bands. This could be understood by considering the symmetry-allowed hybridization between $\beta$ and $\gamma$ bands as they cross each other near $\Gamma$ and hence open up hybridization gap and switch orbital characters. As a result, the $\beta$ band with dominant $d_{yz}$ character acquires $d_{xy}$ character approaching $\Gamma$ while the $\gamma$ band with dominant $d_{xy}$ character becomes $d_{yz}$ like and is degenerate with the $\alpha$ band ($d_{xz}$) above $E_F$ at $\Gamma$. Such conclusions were also reached in another study [26]. We also note that the $d_{xz}$ electron band ($\delta$) bottom is degenerate with the $d_{yz}$ hole band ($\beta$) top at X in the PM state. A summary of the dominant orbital characters of bands near $E_F$ along $\Gamma-X$ are indicated in Fig. 1c. Those for the $\Gamma-Y$ direction are by $C_4$ symmetry derived by an interchange of x and y (*i.e.*, $d_{yz}$ along $\Gamma-X$ becomes $d_{xz}$ along $\Gamma-Y$).

### 3.2. Electronic structure in the SDW state

Across $T_S$ and $T_{SDW}$ in the low temperature orthorhombic SDW state, significant reconstruction of the electronic structure has occurred, as can be seen from both the FS and band dispersions on twinned NaFeAs (Fig. 1d-f). However, part of the complication comes from domain mixing: the observed dispersions have contribution from both $\Gamma-X$ direction in one domain and $\Gamma-Y$ direction in the orthogonal domain. To sort out the intrinsic electronic reconstruction of a single-domain, we measured detwinned crystal.

Figure 2 shows the measured FS and spectral images along $\Gamma-X$ and $\Gamma-Y$ of detwinned crystal in the SDW state. The FS near $\Gamma$ measured in the two geometries do not appear 90° rotated since the polarization used was different with respect to the crystal axes, resulting in different polarization matrix elements that suppress bands of different orbital symmetries in the two measurements. The fact that they are drastically different demonstrates that the crystal was properly detwinned and that the single-domain electronic structure shows broken rotational symmetry. Note that the high symmetry cut of the twinned crystal (Fig. 1e-f) is indeed a mixture of the Γ−X and Γ−Y cuts (Fig. 2e-f). Comparing these two high symmetry cuts, we see an anisotropy of the bands in a similar fashion as observed on detwinned Ba122, namely that the β band near X is shifted up but shifted down near Y. As this band is dominantly $d_{yz}$ near X and $d_{xz}$ near Y, the offset in energy between these two bands reflects broken rotational symmetry in the orbital degree of freedom. For the electron bands, the shifts are less obvious as their intensity is suppressed by polarization matrix elements. However, the larger $k_F$ opening of the δ electron band along Γ−X in the SDW state seen from remnant intensity (Fig. 2e-f) compared to that in the tetragonal state (Fig. 1c) shows that at least the $d_{xz}$ (δ) electron band along Γ−X is also shifted down. The bottom of the $d_{xz}$ band and the top of the $d_{yz}$ band are no longer degenerate at X as in the tetragonal state. Near Γ, orbital anisotropy can be observed on the γ band near $E_F$. The $k_F$ of the $d_{yz}$ portion along Γ-X moves away from Γ, signaling shifting up of the $d_{yz}$ hole band, while the opposite happens to the $d_{xz}$ portion along Γ-Y, resulting in an elongation of the outer hole pocket along the Γ-X direction. Similar elongation also happens to the inner hole pocket (Fig. 2c-d), but in the Γ-Y direction to obey the rotated $d_{xz}/d_{yz}$ character. However, we cannot observe whether the degeneracy of the $d_{yz}$ and $d_{xz}$ band tops is lifted at Γ as they remain above $E_F$. Compared to Ba122 [12], the direction of the orbital anisotropy is the same, namely bands associated with the $d_{xz}$ orbital shift down and bands associated with the $d_{yz}$ orbital shift up.

     Besides modification to the electronic structure due to orbital anisotropy, there are also new features indicating band folding as a result of SDW ordering (Fig. 2f). The most pronouncing one is the folding of the γ band, resulting in a hole-like band crossing near X. Likewise the $d_{xz}$ electron band (δ) is also folded from X to Γ (marked by dotted red line in Fig. 2f), as evident in the hybridization gap it opens with the α hole band, which is also mostly $d_{xz}$ character along Γ-X. Folding via (π, π) vector of other bands may also occur, but not observed as clearly in this dataset likely due to intensity suppression by polarization matrix elements.

     The crossing of δ band and the folded γ band is especially interesting. The remnant intensity of the crossing points shows the two bands to anti-cross at $E_F$ without opening a gap on the Γ-X high symmetry cut, likely due to their opposite orbital symmetries with respect to Γ-X: $d_{xz}$ for δ and mixed $d_{yz}/d_{xy}$ for γ, resulting in a high intensity spot on the FS. Interestingly, from the FS (Fig. 3a), we observe four such high intensity spots. From the measured band dispersions across each of these spots, we see anti-crossing of bands at $E_F$ (Fig. 3b). This must be the result of the folding of hole bands from Γ and electron bands from X, opening up a hybridization gap

everywhere (Fig. 3 cut2) except at the four high symmetry points in the SDW BZ where hybridization is forbidden by the opposite orbital symmetries of the crossing bands. Two interesting implications should be pointed out. Firstly, anti-crossings on $E_F$ at the extrema points of the FS indicate that the folding pockets are well-nested. Secondly, the separation of the crossings are not the same along $\Gamma$-X ($0.42\pi/a$) and $\Gamma$-Y ($0.28\pi/b$), indicating that the good nesting is achieved on the elongated hole pockets, which is a direct result of the orbital anisotropy. We also note that the anti-crossings along $\Gamma$-X are the protected Dirac nodes predicted by theory [27,28]. This is similar to the case of Ba122 where anti-crossings were also observed in both directions. However, the specific energy positions of the crossings appear material-dependent, as those in $BaFe_2As_2$ cross 30meV below $E_F$ along $\Gamma$-Y, indicating varying degree of nesting conditions in different materials depending on the details of the electronic structure.

### 3.3. Temperature evolution of electronic structure

Next, we examine how these electronic modifications evolve from the PM state to the SDW state through a detailed temperature dependence study. In Fig. 4a-b, we show the spectral images taken on the Y-$\Gamma$-X high symmetry lines on a detwinned crystal taken at 10K (T<$T_{SDW}$), 43K ($T_{SDW}$<T<$T_S$), and 70K ($T_S$<T), together with the $\beta$ and $\delta$ band dispersions derived from the tetragonal PM state for comparison. Qualitatively, we see that on one hand, the anisotropic shifts in the $\beta$ and $\delta$ bands along the two directions clearly persist above $T_{SDW}$, and diminish above $T_S$. On the other hand, the signatures of SDW folding—$\gamma$ hole band folding and anti-crossing with the $\delta$ band near X and the hybridization gap of the folded $\delta$ band ($d_{xz}$) with the $\alpha$ hole band ($d_{xz}$)—are clearly observed for T< $T_{SDW}$, but disappears just above $T_{SDW}$. The temperature dependence of these electronic evolutions can be further visualized quantitatively. First, we look at the $\beta$ band shifts near X/Y, and measure its shifting energy position as a function of temperature. This can be done directly for the strain-free twinned crystal as the $\beta$ band from both $\Gamma$-X and $\Gamma$-Y directions appear on the same cut due to domain mixing (Fig. 4c-d). The energy positions of the $\beta$ band from the two directions can be determined simultaneously from the corresponding peaks in the energy distribution curve (EDC) on twinned crystal. We plot the EDC for a representative momentum position A at each temperature, where the peaks marked by red/green diamonds are the $\beta$ bands from $\Gamma$-Y and $\Gamma$-X directions, respectively (Fig. 4e). In addition, a third peak near $E_F$ appears due to $\gamma$ band folding, which will be discussed later. The positions of these peaks can be better determined from the second derivatives of the EDCs, where the band positions show up as dips (Fig. 4f). From the temperature dependence of the energy positions of these two bands (Fig.

4g), we see that they are degenerate in energy above $T_S$, as expected under $C_4$ symmetry in the tetragonal state. Near $T_S$, they begin to separate in energy, signaling the onset of orbital anisotropy already at the structural transition. It is interesting to note that the energy separation already reaches near completion by $T_{SDW}$, below which minimal change is observed. The full energy separation is on the order of 30meV in NaFeAs, compared to 60meV in BaFe$_2$As$_2$, which has both bigger magnetic moment (0.8$\mu_B$ [29] compared to 0.3$\mu_B$ for NaFeAs [30]) and higher transition temperatures ($T_S = T_{SDW} = 140$K). The same temperature dependent measurements can be done on the orthogonal directions on detwinned crystals, which are plotted in Fig. 4g for comparison. The major difference is that the anisotropy on detwinned crystals onsets far above $T_S$ (defined for unstressed crystals), similar to the observation in detwinned Ba122. This is consistent with detwinning pressure dependent resistivity measurement [6] showing smearing of $T_S$ due to the symmetry-breaking field of the uniaxial strain, suggesting a large nematic susceptibility and strong orbital fluctuations associated with $T_S$ as revealed by the detwinning stress.

Next, we look at the signatures of band folding by examining the temperature dependence of the folded $\gamma$ hole band near X from both EDC and MDC. On the EDC, as seen from twinned sample (Fig. 4e), the peak closest to $E_F$ (marked in blue) corresponding to the folded $\gamma$ band only appears below $T_{SDW}$. The behavior is even clearer on the MDC taken slightly below $E_F$ on detwinned crystal (Fig. 4h). Two peaks are observed there, corresponding to the folded $\gamma$ band on the left and the original $\delta$ electron band on the right. The feature corresponding to the folded $\gamma$ band only appears below $T_{SDW}$.

To summarize the temperature evolution of electronic reconstruction (Fig. 5), we observe distinct electronic reconstruction onsetting at $T_S$ and $T_{SDW}$. For strain-free twinned crystals, rotational symmetry breaking modification to the electronic structure begins near $T_S$ as orbital anisotropy onsets, manifested in the upward shifting of $d_{yz}$ bands and downward shifting of $d_{xz}$ bands. Such anisotropy develops towards $T_{SDW}$, at which temperature long range magnetic ordering sets in and band folding occurs on top of the anisotropic electronic structure due to the broken translational symmetry, resulting in smaller reconstructed FS pockets. From the experimental point of view, measurement on detwinned crystal is crucial for disentangling the electronic reconstruction and symmetry breaking process through each transition.

## 4. Discussion

One of the current focuses in the pnictide community is to understand the dominating mechanisms responsible for the pervasive structural and magnetic transitions in the phase diagram and their connection to the emerging superconductivity. Many theoretical proposals have been put forth in this regard, especially in light of many recent experimental findings of the large electronic nematicity associated with the structural transition. Such theoretical endeavors include those driven by magnetism [31-35], in which a symmetry-breaking Ising nematic phase can emerge above $T_{SDW}$ via spin fluctuations; those based on orbital ordering [36-42] reflected in symmetry breaking via an unequal occupation of the $d_{yz}$ and $d_{xz}$ orbitals onsetting at the higher symmetry-breaking phase transition of $T_S$; those based on a Pomeranchuk instability between the near nesting of the $C_4$ symmetric $d_{xy}$ hole pocket and the elliptical electron pockets [43]; and symmetry breaking from a valley-density wave and hidden orders [44]. Experimental understanding of what occurs at these two transitions is important to distinguish amongst these proposals.

Here in this work, the observed sequential development of orbital anisotropy at the higher $T_S$ and then band folding at the lower $T_{SDW}$ reveals an interesting connection between these two transitions. As we have noted, the electron and hole bands that fold appear to have good nesting condition. However, such nesting is observed to be achieved at $T_{SDW}$ *after* the full development of anisotropic band shift between $T_S$ and $T_{SDW}$, evident in the elongated shape of the folding hole pockets. One natural interpretation of this observation is that at $T_S$, orbital anisotropy begins to develop, breaking the $C_4$ rotational symmetry of the FS and lifting the degeneracy of the nesting instabilities between $\Gamma$-X and $\Gamma$-Y as one is enhanced and the other suppressed. At $T_{SDW}$, the nesting instability of the favored direction diverges and drives the collinear AFM order as the bands fold in this direction. Such a scenario was also proposed by Kontani *et al.* based on the orbital fluctuation theory [45]. This sequence of electronic evolution also explains the observed onset of an enhancement of spin fluctuations at $T_S$ and its divergence towards $T_{SDW}$ in NMR measurements of NaFeAs [30].

In a bigger context, we see that the case of NaFeAs and that of Ba122 [12] are very similar: i) the anisotropic orbital/band shift is always associated with the rotational symmetry breaking at $T_S$; ii) the direction of orbital anisotropy is the same with $d_{xz}$ ($d_{yz}$) bands shifting down (up). The well-separated transitions and lack of surface complications in NaFeAs here allowed us to see more clearly what occurs at $T_{SDW}$. Such an orbital anisotropy-enhanced nesting-driven AFM order implies that the underlying electronic structure with its orbital content plays a governing role for these two transitions. While the large orbital anisotropy of the nematic phase may be a general phenomenon for iron pnictides that share similar electronic structure, the subtle difference in the electronic structure may be held accountable for the variation of the relation between these two transitions from family to family or with doping, as to be understood with further work. Nonetheless, our observation reveals a strong coupling of the orbital fluctuations associated with the onset of orbital anisotropy at $T_S$ and the enhanced spin fluctuations in the electronic nematic phase. And the consideration of the interplay between these two fluctuations may be critical in understanding the emerging superconducting phase in the pnictides.

**Acknowledgements**

The authors are grateful for helpful discussions with F. Wang, D.-H. Lee, B. Moritz, A. Sorini, A. Kemper, C.-C. Chen, Y. Zhang and D. L. Feng. ARPES experiments were performed at the Stanford Synchrotron Radiation Lightsource, operated by the Office of Basic Energy Science, U.S. DOE. The Stanford work is supported by DOE Office of Basic Energy Science, Division of Materials Science and Engineering, under contract DE-AC02-76SF00515. MY thanks the NSF Graduate Research Fellowship Program for financial support.

*Electronic reconstruction in detwinned NaFeAs*

*Electronic reconstruction in detwinned NaFeAs*

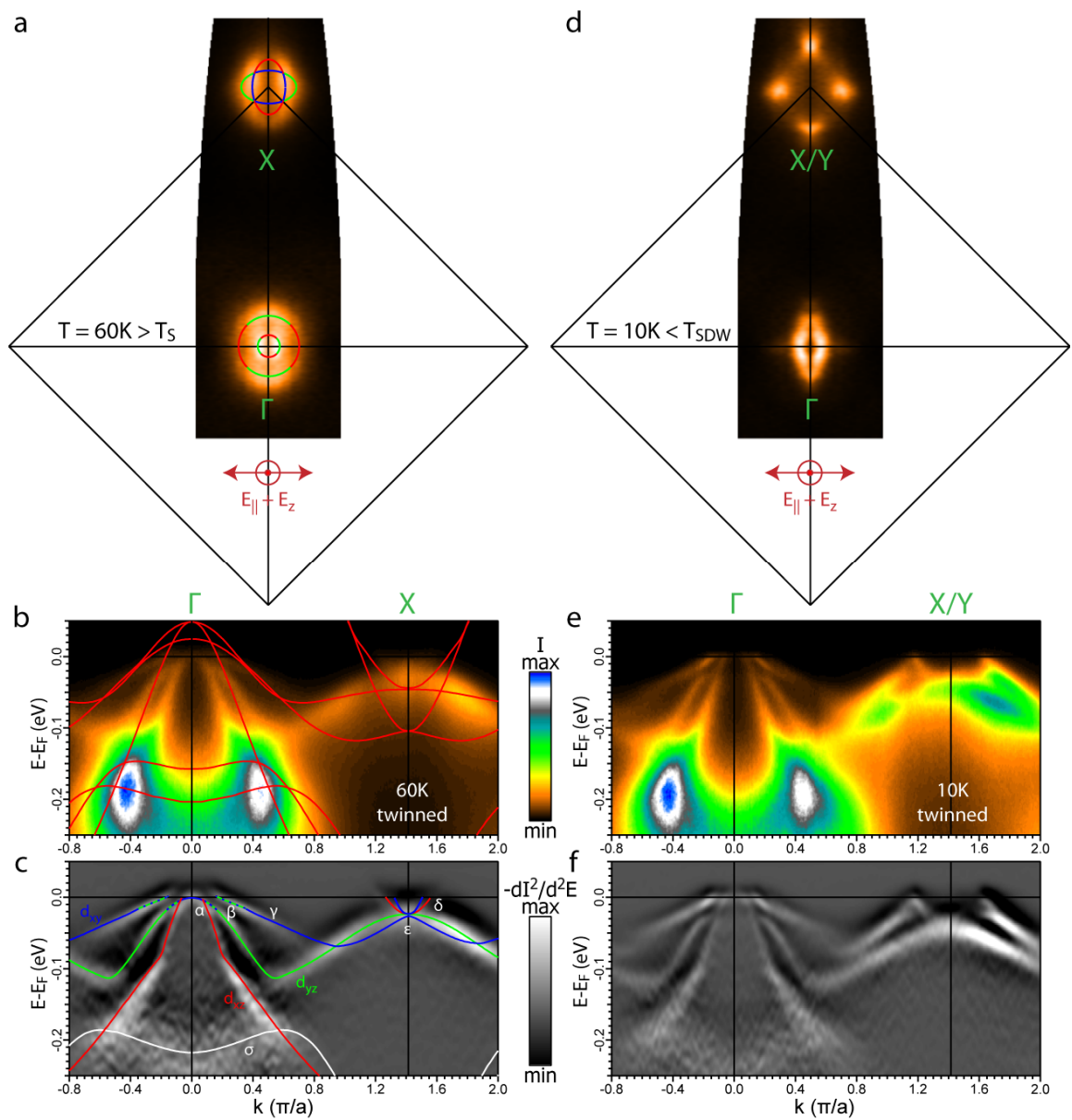

**Figure 1**

**Electronic structure of twinned NaFeAs**. (a) FS measured in the tetragonal PM state at 60K. (b) Spectral image along the Γ-X high symmetry cut, overlaid with LDA calculations which were renormalized by a factor of 4. (c) Second derivative of (b), overlaid with band guides to eye marked with dominant orbital contribution (red/ green/blue: $d_{xz}/d_{yz}/d_{xy}$). Dotted colors denote mixed orbital characters (see text). (d)-(f) Same as that of (a)-(c) but taken in the SDW state at 10K. The FS were symmetrized according to crystal symmetry.

*Electronic reconstruction in detwinned NaFeAs*

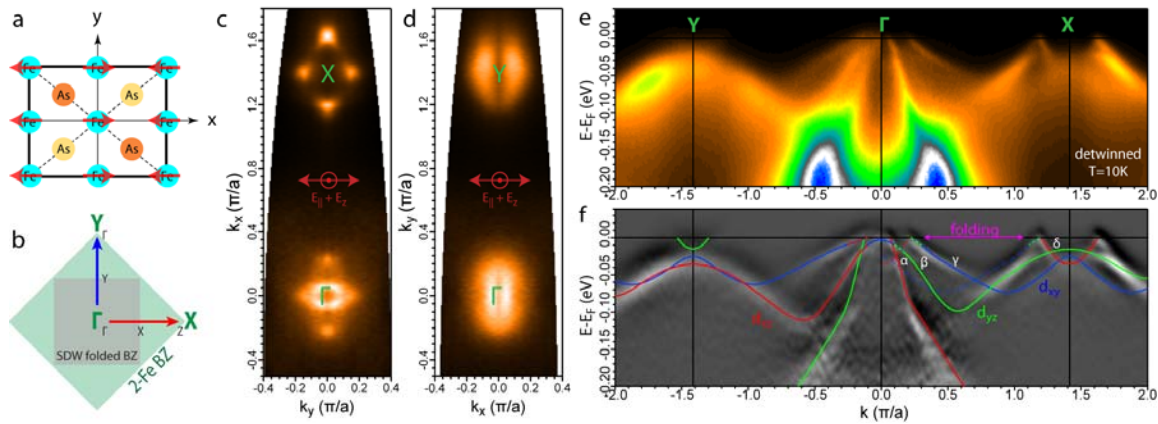

**Figure 2**

**Electronic anisotropy observed in the SDW state of detwinned NaFeAs**. (a) Schematic of orthorhombic SDW unit cell. (b) BZ definitions for the tetragonal PM state (green) and orthorhombic SDW state (gray). PM BZ (green) notations are used throughout, where Γ-X (Γ-Y) is along the AFM (FM) direction. (c)-(d) FS measured along Γ-X and Γ-Y directions, respectively, with polarization vectors labeled in red. (e) Spectral image along Y-Γ-X high symmetry directions. (f) Second derivative image of (e). Band eye guides for prominent observable bands are marked with dominant orbital characters. Weaker dotted lines denote folded bands. All data were measured at 10K on detwinned crystal.

*Electronic reconstruction in detwinned NaFeAs*

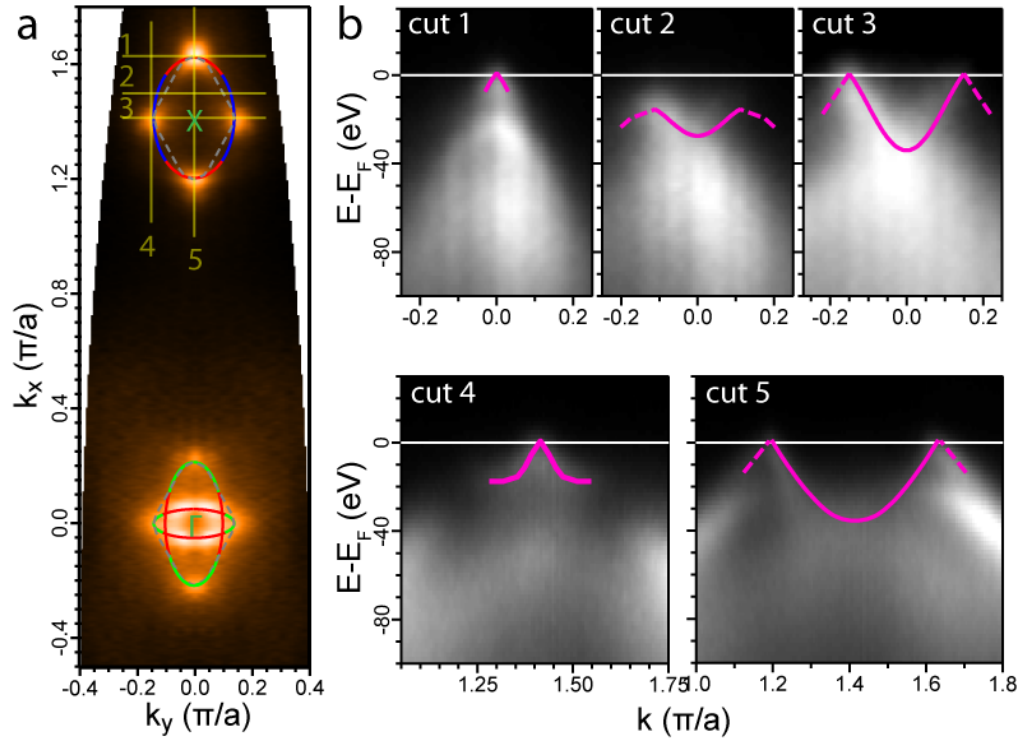

**Figure 3**

**Band crossings of folded hole and electron bands in the SDW state**. (a) FS measured along Γ-X on detwinned crystal, where the folding hole pockets at Γ (blue) and electron pocket at X are marked. Dotted contour mark where hybridization gap is opened. (b) Spectral images along various cuts marked in (a), where zero gap is observed at the high symmetry points of folded bands, and finite hybridization gap is observed elsewhere. Measurements taken at 10K.

*Electronic reconstruction in detwinned NaFeAs*

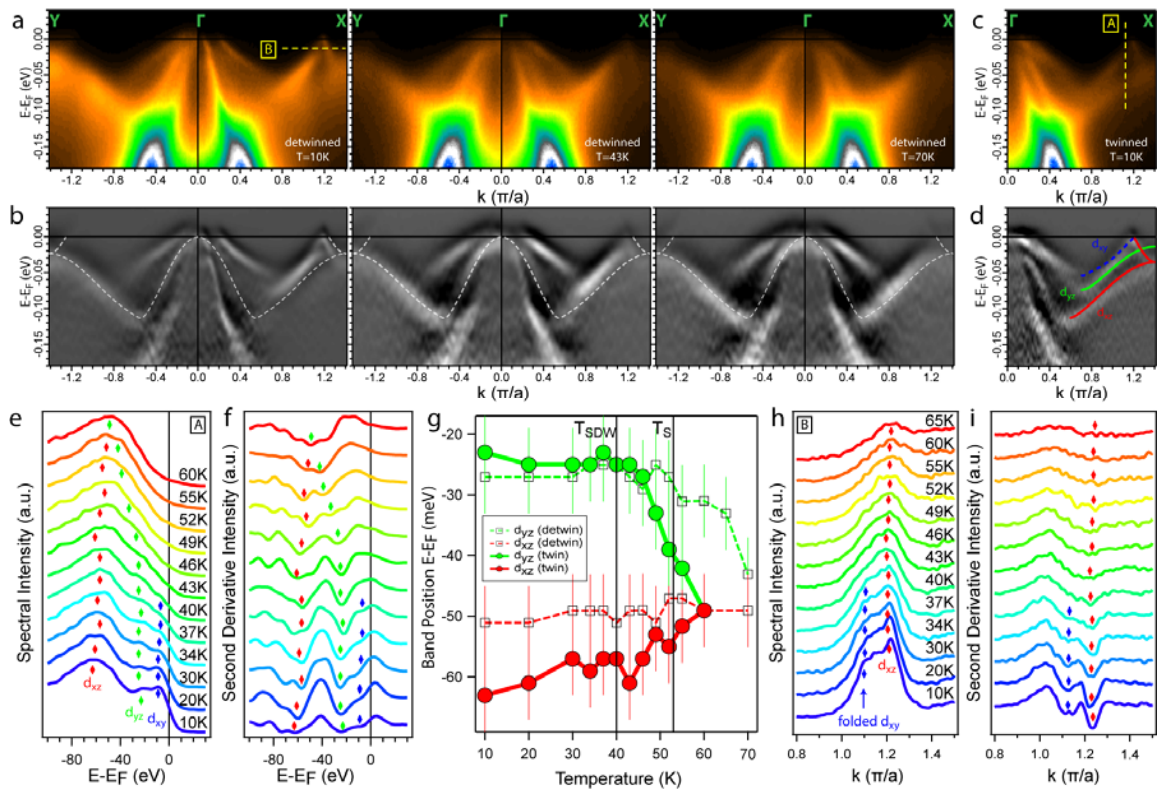

**Figure 4**

**Temperature evolution of the electronic reconstruction**. (a)-(b) Spectral image and its second derivative along Γ-X/Y high symmetry direction measured at 10K on twinned crystal. (c)-(d) Spectral image and its second derivative along Y-Γ-X high symmetry cuts on detwinned crystal at 10K (T<$T_{SDW}$), 43K ($T_{SDW}$<T<$T_S$), and 70K ($T_S$<T). Dotted lines mark the β band position in the PM state for comparison. (e) Temperature evolution of the EDC cut labeled as A in (a), where peak positions are marked to indicate band positions. (f) Second derivatives of the EDCs in (e), where dips indicate band positions. (g) Temperature dependence of the $d_{yz}$ (green) and $d_{xz}$ (red) band positions measured at momentum A on twinned and detwinned crystals. (h) Temperature evolution of the MDC cut labeled as B in (c). (i) Second derivatives for the MDCs in (h).

*Electronic reconstruction in detwinned NaFeAs*

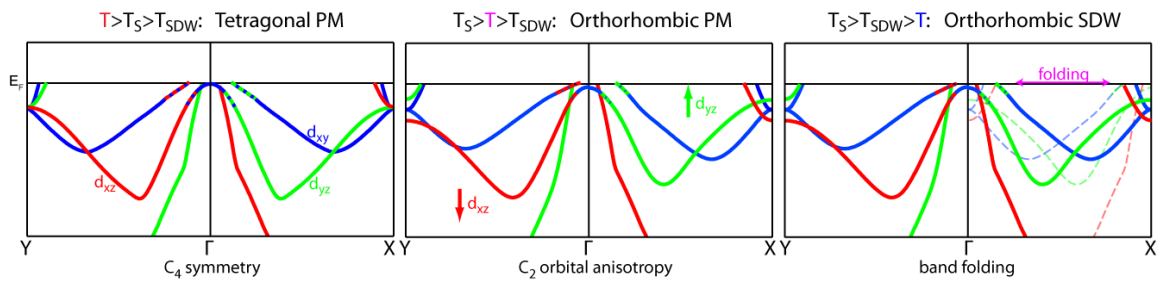

**Figure 5**

**Schematic of electronic structure evolution**. In the tetragonal PM state, the electronic structure is $C_4$ symmetric. As temperature is cooled upon $T_S$, $C_4$ symmetry is broken as bands of dominant $d_{xz}$ character shift down and bands of dominant $d_{yz}$ character shift up. Upon further cooling, at $T_{SDW}$, band folding occurs along the AFM direction ($\Gamma$-X), and hybridization gaps may open where allowed by orbital symmetry (not shown).